\documentclass[a4paper,11pt]{article}

\usepackage{jcappub} 

\usepackage[T1]{fontenc} 

\usepackage{hyperref}
\hypersetup{
  colorlinks=true,        
  linkcolor=blue,         
  citecolor=magenta,         
}

\def\beq{\begin{equation}}
\def\eeq{\end{equation}}
\def\bear{\begin{eqnarray}}
\def\ear{\end{eqnarray}}

\title{\boldmath Gravitational instability of polytropic spheres containing region of trapped null geodesics: a possible explanation of central supermassive black holes in galactic halos}

\author[a]{Zden\v{e}k Stuchl\'{i}k,}
\author[a]{Jan Schee,}
\author[a,b,1]{Bobir Toshmatov,\note{Corresponding author.}}
\author[a]{Jan Hlad\'{i}k,}
\author[a]{and Jan Novotn\'{y}}

\affiliation[a]{Institute of Physics and Research Centre of Theoretical Physics and Astrophysics,\\ Faculty of Philosophy \& Science, Silesian University in Opava,\\ Bezru\v{c}ovo n\'{a}m\v{e}st\'{i} 13, CZ-74601 Opava, Czech Republic}
\affiliation[b]{Institute of Nuclear Physics,\\ Ulughbek, Tashkent 100214,
Uzbekistan}

\emailAdd{zdenek.stuchlik@fpf.slu.cz}
\emailAdd{jan.schee@fpf.slu.cz}
\emailAdd{bobir.toshmatov@fpf.slu.cz}
\emailAdd{jan.hladik@fpf.slu.cz}
\emailAdd{jan.novotny@fpf.slu.cz}

\abstract{We study behaviour of gravitational waves in the recently introduced general relativistic polytropic spheres containing a region of trapped null geodesics extended around radius of the stable null circular geodesic that can exist for the polytropic index $N>2.138$ and the relativistic parameter, giving ratio of the central pressure $p_\mathrm{c}$ to the central energy density $\rho_\mathrm{c}$, higher than $\sigma = 0.677$. In the trapping zones of such polytropes, the effective potential of the axial gravitational wave perturbations resembles those related to the ultracompact uniform density objects, giving thus similar long-lived axial gravitational modes. These long-lived linear perturbations are related to the stable circular null geodesic and due to additional non-linear phenomena could lead to conversion of the trapping zone to a black hole. We give in the eikonal limit examples of the long-lived gravitational modes, their oscillatory frequencies and slow damping rates, for the trapping zones of the polytropes with $N \in (2.138,4)$. However, in the trapping polytropes the long-lived damped modes exist only for very large values of the multipole number $\ell>50$, while for smaller values of $\ell$ the numerical calculations indicate existence of fast growing unstable axial gravitational modes. We demonstrate that for polytropes with $N \geq 3.78$, the trapping region is by many orders smaller than extension of the polytrope, and the mass contained in the trapping zone is about $10^{-3}$ of the total mass of the polytrope. Therefore, the gravitational instability of such trapping zones could serve as a model explaining creation of central supermassive black holes in galactic halos or galaxy clusters.}

\arxivnumber{1704.07713}

\begin{document}
\maketitle
\flushbottom

\section{Introduction}\label{sec-intr}

Extremely compact (or ultra-compact) objects, defined in the standard way as objects with surface $R$ located under the radius $r_\mathrm{ph}$ of photon circular geodesic of the external Schwarzschild (or a generalized spherically symmetric) vacuum spacetime, are astrophysically very interesting because they have to contain a region of trapped null geodesics that could be relevant for trapping of gravitational waves \cite{1999A-GRO}, or radiated neutrinos \cite{2012SHU-NTE}. Existence of the extremely compact objects has been demonstrated in the principally very interesting case of spheres with uniform distribution of energy density, but radius dependent pressure \cite{2000S-SSS,2001SHS-NGE,2004B-ESS,2011HS-PNR}. Unfortunately, the neutron or quark star models based on the known realistic equations of state do not allow for the existence of extremely compact objects defined in this way, requiring in the most compact models (in the geometric units with $c=G=1$) $R \geq 3.5M > r_\mathrm{ph} = 3M$ where $M$ denotes mass of the compact star \cite{2006LP-ESN}.

Recent study of the general relativistic polytropic spheres in spacetimes with a repulsive cosmological constant demonstrates possibility to obtain polytropic spheres containing near their center a region with trapped null geodesics~\cite{2016SHN-GRP}. This is an important and surprising result as surface of such polytropes can be located highly above the $r_\mathrm{ph}=3M$ surface, so that we could reconsider definition of the extremely compact objects restricting attention solely to the existence of the zone of the trapped null geodesics. We thus introduce the notion of trapping polytropes. \footnote{We have to stress that in the trapping polytropes the compactness condition that is relevant for the ultra-compact objects, is not satisfied even locally, as demonstrated explicitly in \cite{2017NHS}.} Their trapping zone is governed by two photon circular geodesics -- the inner one at $r_{ph(s)}$ that is stable relative to radial perturbations, and the outer one at $r_{ph(u)}$ that is unstable. The radius $r_{ph(u)}$ can be considered as the boundary of the trapping zone, and in the trapping polytropic spheres with sufficiently high values of the parameters $N$ and $\sigma$, it can be by many orders smaller than the polytrope radius.

The polytropic spheres represent useful physical idealization of behaviour of fluid configurations in various conditions. It is well know that they represent both the non-relativistic and ultra-relativistic degenerated Fermi gas that can be taken quite seriously, being relevant and physically interesting especially for the ultra-relativistic Fermi gas applied as basic approximation of matter in the inner core region of neutron stars~\cite{1983ST-BHW}; however, the polytropic equations of state are also used to approximate the realistic equations of state governing structure of neutron stars~\cite{2009Ozel-Psaltis}. On the other hand, the extremely extended polytropic structures can model also dark-matter halos of galaxies or galaxy clusters -- surprisingly, both non-relativistic and relativistic polytropic spheres could be relevant~\cite{2016SHN-GRP,2017NHS}. As there is no clear candidate of the particle physics for explanation of the dark matter, we are free in choice of the polytrope equation of state, and only the observational restrictions coming from the observations of the velocity curves in galaxies and galaxy clusters could serve as a guide for giving restrictions on the character of the polytropic equation of state governing matter in halos. In a recent study of general relativistic polytropes, including the effect of the observationally limited cosmological constant, it has been demonstrated that the relativistic trapping polytropes can have their extension and mass comparable with the extension and mass of the large galaxies and galaxy clusters; moreover, the orbital velocity of the circular geodesics of such relativistic trapping polytrope spacetimes can be matched to the velocity profiles observed in the outer regions of large galaxies and in galaxy clusters~\cite{Stu-Nov-Hle:submitted2017}.

In detailed study of the existence of general relativistic trapping polytropes, related to the models of compact stars~\cite{2017NHS}, the role of the cosmological constant has been abandoned, as it is related only to very extended objects with radius close to the static radius of the external spacetime~\cite{1983S-MTP,1999SH-SPS,2005S-IRC} and central density much smaller than the nuclear density~\cite{2016SHN-GRP}. It has been shown that the trapping polytropes must have the polytropic index $N > 2.138$ and relatively large relativistic parameter $\sigma > 0.677$. In the $N - \sigma$ parameter space, extension of the trapping region increases as the polytropic index $N$ increases, being restricted from above by the value of $\sigma$ corresponding to the causal limit~\cite{2017NHS}. In order to relate the trapping phenomenon to astrophysically relevant objects, namely the neutron (quark) stars, validity of the polytropic configurations has been restricted in~\cite{2017NHS} to their extension $r_\mathrm{extr}$ corresponding to the gravitational mass $M \sim 2M_{\odot}$ of the most massive observed neutron stars. For the central density $\rho_{c} \sim 5 \times 10^{15}$~g\,cm$^{-3}$ the whole trapped regions are located inside of $r_\mathrm{extr}$ (i.e., there is $r_\mathrm{extr} > r_{ph(u)}$) for $2.138 < N < 3.1$, while for $\rho_\mathrm{c} \sim 10^{16}$~g\,cm$^{-3}$, the whole trapped regions are inside of $r_\mathrm{extr}$ for values of $2.138 < N < 4$, guaranteeing astrophysically plausible trapping for all considered polytropes in the whole region limited by the radius $r_{ph(u)}$. In the case $r_\mathrm{extr} < r_{ph(u)}$, the trapping zone is limited by the radius $r_\mathrm{extr}$; above this radius the vacuum Schwarzschild geometry is relevant. In the present paper, we consider whole the trapping polytropic spheres.

The effective potential of the null geodesics of the trapping polytrope internal spacetime has a local maximum (minimum) corresponding to the stable (unstable) circular null geodesic; the trapping zone is concentrated around the stable circular geodesics, being restricted from above by the unstable circular geodesic \cite{2017NHS}. This effective potential resembles the related effective potential in the ultra-compact uniform density objects \cite{Kokkotas:MNRAS:1992,Kojima:PTP:1988,Kokkotas:MNRAS:1994,Allen:PRD:1998,Cardoso:PRD:2014}, and in the Kerr naked singularity spacetimes~\cite{Stuchlik:CQG:2010:}; at the region located above the stable circular null geodesic, it is similar also to the effective potential of the black hole spacetimes~\cite{1970Zerilli}. Considering the behaviour of the gravitational wave perturbations of the trapping polytrope spacetimes, we thus expect similar phenomena as those predicted in the spherically symmetric black hole~\cite{1970Zerilli} and, especially, in the standard spherically symmetric ultra-compact uniform density objects where the so called long-lived axial gravitational perturbations with extremely small damping occur that can cause in an additional non-linear regime a gravitational instability collapse of the trapping zone~\cite{Pani:PRD:2009,Pani:PRD:2010,Cardoso:PRD:2014,Macedo:IJMPD:2015}. Note that in the Kerr naked singularity spacetimes unstable, i.e., growing axial gravitational modes occur~\cite{Cardoso:PRD:2006}, and similarly, in the rotating ultra-compact stars the long-lived modes can be turned to unstable growing modes too, due to the linear ergoregion instability \cite{Cardoso:PRD:2014}.

We thus study in the present paper the possible existence of the gravitational instability related to the so called long-lived axial gravitational quasi-normal modes, and its astrophysical consequences. We first summarize solutions of the general relativistic polytropic spheres and the special class of the trapping polytropes. As the metric coefficients of the polytropic spheres are determined numerically only, the evolution equation of the linear gravitational perturbations has to be given in the general form in the second part of the paper. Then we look for the long-lived modes of the gravitational perturbative solutions studying the wave equation. We concentrate out attention to the axial gravitational perturbations located in the trapping region concentrated around the stable circular null geodesic. The numerical calculations show that along with the long-lived damping modes, also fast growing unstable axial modes exist. Finally, we study the relation of the extension of the trapping zone to the extension of the whole polytrope, and the relation of the amount of the polytrope mass contained in the trapping zone that could form a black hole due to a collapse ignited by a non-linear effect related to the long life of the linear axial gravitational modes, as proposed in \cite{Cardoso:PRD:2014,Macedo:IJMPD:2015}, to the total mass of the polytrope. In the present study we simplify the problem by considering vanishing cosmological constant, as it can be expected that the estimates on the long-lived gravitational modes related to the trapping zones, given under this simplification, can be precise enough due to the fact that the trapping zone is always very close to the centre of the configuration, where the cosmic repulsion can be abandoned.


\section{General relativistic polytropes}\label{eost}

In terms of the standard Schwarzschild coordinates, the line element of a spherically symmetric, static spacetime of a polytropic sphere takes the form
\beq
    \mathrm{d}s^{2} = - f(r) c^{2} \mathrm{d}t^{2} + \frac{1}{g(r)} \mathrm{d}r^{2} + r^{2} (\mathrm{d}\theta^{2} + \sin^{2} \theta \mathrm{d}\phi^{2}),
\eeq
with two unknown metric functions depending only on the radial coordinate, $f(r)$ and $g(r)^{-1}$. The static configuration is assumed to be constituted from a perfect fluid with the stress-energy tensor
\beq
    T^\mu_{\hphantom{\mu}\nu} = (p+\rho c^{2}) U^{\mu} U_{\nu} + p\,\delta^{\mu}_{\nu},
\eeq
where $U^{\mu}$ denotes the 4-velocity of the fluid. In the fluid rest-frame we denote by $\rho = \rho (r)$  the mass-energy density and by $p = p(r)$ the isotropic pressure that are related by the polytropic equation of state
\beq
    p = K \rho^{1+\frac{1}{N}},
\eeq
where constant $N$ denotes the polytropic index. $K$ denotes a constant governed by the thermal characteristics of a given polytrope by specifying the density $\rho_{\mathrm{c}}$ and pressure $p_{\mathrm{c}}$ at its center \cite{1964T-GRP}
\beq
    \sigma  \equiv \frac{p_{\mathrm{c}}}{\rho_{\mathrm{c}} c^{2}}\ = \frac{K}{c^{2}}\rho_{\mathrm{c}}^{\frac{1}{N}}.
\eeq
The constant $K$ contains the temperature implicitly, as for a given pressure, the density is a function of temperature. The polytropic equation represents the parametric equations of state for the completely degenerate gas at zero temperature, relevant, e.g., for neutron stars, when both $N$ and $K$ are universal physical constants \cite{1964T-GRP}. The equation of state of the ultra-relativistic degenerate Fermi gas is determined by the polytropic equation with the polytropic index $N = 3$, while the non-relativistic degenerate Fermi gas $N = 3/2$ \cite{1983ST-BHW}.

\subsection{Polytrope structure equations}
The structure equations of the general relativistic polytropic spheres are given by the Einstein equations
\beq
    R_{\mu\nu} - \frac{1}{2}Rg_{\mu\nu} = \frac{8\pi G}{c^4} T_{\mu\nu},
\eeq
and the energy-momentum conservation law
\beq
    T^{\mu\nu}_{\hphantom{\mu\nu};\nu} = 0.
\eeq

The polytrope structure is governed by the two structure functions. The first one, $\theta(r)$, gives the energy density radial profile $\rho(r)$ through the central density $\rho_{\mathrm{c}}$ by the relation \cite{1964T-GRP}
\beq
    \rho = \rho_{\mathrm{c}} \theta^{N},
\eeq
with the boundary condition $\theta(r=0)=1$. The second one is the mass function
\beq
    m(r) =  \int^{r}_{0} {4 \pi r^{2} \rho \mathrm{d}r},
\eeq
fixed by the integration constant chosen to be $m(0) = 0$ that guarantees smooth spacetime geometry at the origin \cite{1973MTW-G}. The surface of the polytrope is at $r=R$, where $\rho(R)=p(R)=0$, and its total mass reads $M=m(R)$. Exterior of the polytrope is described by the vacuum Schwarzschild geometry.

Introducing the characteristic length scale $\mathcal{L}$ of the polytrope \cite{1964T-GRP}
\beq
    \mathcal{L} = \left[\frac{(N+1)K\rho_{\mathrm{c}}^{1/N}}{4\pi G\rho_{\mathrm{c}}}\right]^{1/2} = \left[\frac{\sigma(N+1)c^{2}}{4\pi G\rho_{\mathrm{c}}}\right]^{1/2},
\eeq
and its characteristic mass scale $\mathcal{M}$
\beq
    \mathcal{M} = 4\pi \mathcal{L}^3 \rho_c = \frac{c^2}{G}\sigma (N+1)\mathcal{L},
\eeq
dimensionless radial coordinate
\beq
    \xi = \frac{r}{\mathcal{L}},
\eeq
and dimensionless gravitational mass function
\begin{align}
    v(\xi) &= \frac{m(r)}{4\pi \mathcal{L}^{3}\rho_{\mathrm{c}}},
\end{align}
the dimensionless structure equations related to the two structure functions, $\theta(r)$ and $m(r)$, and the parameters $N$, $\sigma$, can be put into the dimensionless form \cite{1964T-GRP,2016SHN-GRP,2017NHS}
\begin{align}
    \xi^{2}\frac{\mathrm{d}\theta}{\mathrm{d}\xi}\frac{1-2\sigma(N+1)\left(v/\xi\right)}{1+\sigma\theta} + v(\xi)&= - \sigma\xi\theta\frac{\mathrm{d}v}{\mathrm{d}\xi}, \label{grp31}\\
    \frac{\mathrm{d}v}{\mathrm{d}\xi} &= \xi^{2}\theta^{N}. \label{grp32}
\end{align}

For fixed parameters $N$, $\sigma$, the structure equations (\ref{grp31}) and (\ref{grp32}) can be  simultaneously solved with the boundary conditions,
\beq
\theta(0) = 1, \quad v(0) = 0.    \label{grp33}
\eeq
Using Eqs.~(\ref{grp32}) and (\ref{grp33}) we find that $v(\xi) \sim \xi^{3}$ for $\xi \to 0$ and, according to Eq.\,(\ref{grp31}), we obtain
\beq
\lim_{\xi\to 0_+}\frac{\mathrm{d}\theta}{\mathrm{d}\xi} = 0.
\eeq
The surface of the polytropic sphere at $r = R$, is given by the first zero point of $\theta(\xi)$ that is denoted as $\xi_{1}$:
\beq
\theta(\xi_{1}) = 0.
\eeq
Then the solution $v(\xi_{1})$ determines the polytrope gravitational mass.

\subsection{Characteristics of the polytropic spheres}
For fixed parameters $N$, $\sigma$ and $\rho_c$ of the polytrope we directly obtain the scale factors $L$, $\mathcal{M}$, and numerical calculations give the two solutions of the structure equations $\xi_{1}$ and $v(\xi_{1})$ determining all the characteristics of the polytrope. The radius of the polytropic sphere reads
\beq
R = \mathcal{L} \xi_{1},
\eeq
its gravitational mass is given by
\beq
M = \mathcal{M} v(\xi_1) = \frac{c^{2}}{G} \mathcal{L}\sigma(N+1) v(\xi_{1}),
\eeq
and the radial profiles of the energy density, pressure, and mass-distribution are determined by the relations
\bear
\rho(\xi) &=& \rho_{\mathrm{c}}\theta^{N}(\xi),\\
p(\xi) &=& \sigma\rho_{\mathrm{c}}\theta^{N+1}(\xi),\\
M(\xi) &=& M\frac{v(\xi)}{v(\xi_{1})}.
\ear
The radial structure of the polytropic spheres is thus always fully determined by the parameters $N$ and $\sigma$ only, while their extension and mass are determined also by the central density parameter $\rho_c$.~\footnote{If the influence of the cosmological constant is included in the determination of the polytropic spheres, we find that also the radial structure of the polytropic spheres is governed by the central density, being mixed with the vacuum density related to the cosmological constant \cite{2016SHN-GRP}. The cosmological constant thus breaks the degeneracy in the radial structure of polytropic spheres.} Detailed discussion of the polytrope properties, including their gravitational binding energy and the internal energy, can be found in \cite{2016SHN-GRP,1964T-GRP}. Effectiveness of the gravitational binding of the polytrope is given by the compactness parameter defined by the relation
\beq
\mathcal{C} \equiv \frac{GM}{c^{2}R} = \frac{1}{2}\frac{r_{\mathrm{g}}}{R} = \frac{\sigma(N+1)v(\xi_{1})}{\xi_{1}},
\eeq
where the standard gravitational radius of the polytrope, reflecting its gravitational mass in length units,
\beq
r_{\mathrm{g}} = \frac{2GM}{c^{2}}
\eeq
is used. The compactness parameter $\mathcal{C}$ can be reflected by the gravitational redshift of radiation emitted from the surface of the polytrope~\cite{2011HS-PNR}.

All the characteristic functions of the polytrope interior can be determined only by numerical procedures for the polytropes with $N>0$~\cite{1964T-GRP,2016SHN-GRP}. The special case of $N=0$ corresponds to physically unrealistic polytropic configurations with uniform distribution of energy density when the characteristic functions can be given in terms of elementary functions; the $N=0$ polytropes could thus serve as a test bed for more complex general polytropes~\cite{2000S-SSS,2016SHN-GRP}.

The exterior of the polytropic sphere is represented by the vacuum Schwarzschild spacetime with the same gravitational mass parameter $M$ as those characterizing the internal spacetime of the polytrope, and is given by the metric coefficients
\beq
f(r) = g(r) = 1 - \frac{2GM}{c^{2} r}.
\eeq
Note that in the astrophysically relevant constructions of the neutron stars, the polytropic equation of state can be applied only up to the radius giving the realistic mass of the polytropic sphere, denoted as $r_{extr}$~\cite{2017NHS}. Then the polytrope model represents the central part of the neutron star corresponding to its inner core \cite{1983ST-BHW}; even sequence of several polytropic equations of state can be applied as shown in~\cite{2009Ozel-Psaltis}.~\footnote{The most precise modeling is based on application of the density criterion, as used in the standard models of neutron stars~\cite{1983ST-BHW,1999G}.} In the present study concentrated on the behaviour of the gravitational waves, we use the single polytropes with their complete extension.

The photon sphere of the vacuum Schwarzschild spacetime, given by the photon circular geodesics, is located at the radius~\cite{1973MTW-G}
\beq
r_\mathrm{ph} = \frac{3GM}{c^{2}} = \frac{3}{2}r_\mathrm{g}.
\eeq
For original definition of the extremely compact objects, the condition $R<r_\mathrm{ph}$ must be satisfied that can be expressed in the form $C > 1/3$. Such extremely compact objects can exist for the special case of the $N=0$ polytropes; the limiting case $R=9r_{g}/8$ can be slightly modified by the presence of the cosmological constant \cite{2000S,2004B-ESS}, but such objects are excluded by realistic equations of state applied in the models of neutron stars \cite{2013UMS}.

For a deeper insight into the character of the polytropes, we can define and study also the locally defined compactness of the polytrope, given by the relation
\beq
    \mathcal{C(\xi)} \equiv \frac{\sigma(N+1)v(\xi)}{\xi}.
\eeq
However, detailed study of the trapping polytropes demonstrated that the condition $C(\xi) > 1/3$ is not satisfied in their interior~\cite{2017NHS}.

Recent study of the polytropic stars with equation of state relating the pressure and the rest-energy density~\cite{Saida:2016} demonstrates existence of configurations with global compactness $C>1/3$, i.e., with surface located under the Schwarzschild photon sphere; however, such configurations have according to the numerical models speed of sound larger than the speed of light. On the other hand, in our polytrope models based on equation of state relating the pressure and the energy density, the considered trapping polytropes have the light speed exceeding the sound speed -- for details see~\cite{2017NHS}.

\section{General relativistic trapping polytropes and their circular null geodesics}

Numerical calculations of the structure equations of the polytrope spheres yield the dimensionless radial profiles of energy density, mass and metric coefficients, and the dimensionless extension and mass parameters $\xi_1$ and $v_1 = v(\xi_1)$ in dependence on the polytropic index $N$ and the relativistic parameter $\sigma$. These polytrope characteristics are independent of the third parameter, $\rho_\mathrm{c}$, that governs the length and mass scales of the polytropes. Therefore, it is convenient to study the trapping polytropes in the dimensionless formalism, simplifying significantly the discussion of the properties of the polytropic spheres. The trapping polytropes were found and thoroughly discussed in \cite{2017NHS}. Here we present a short summary related to the properties of the null geodesics of the internal polytrope spacetime.

\subsection{Geometry of the internal polytrope spacetime}

\begin{figure*}[tbp]
\begin{center}
\includegraphics[width=0.47\linewidth]{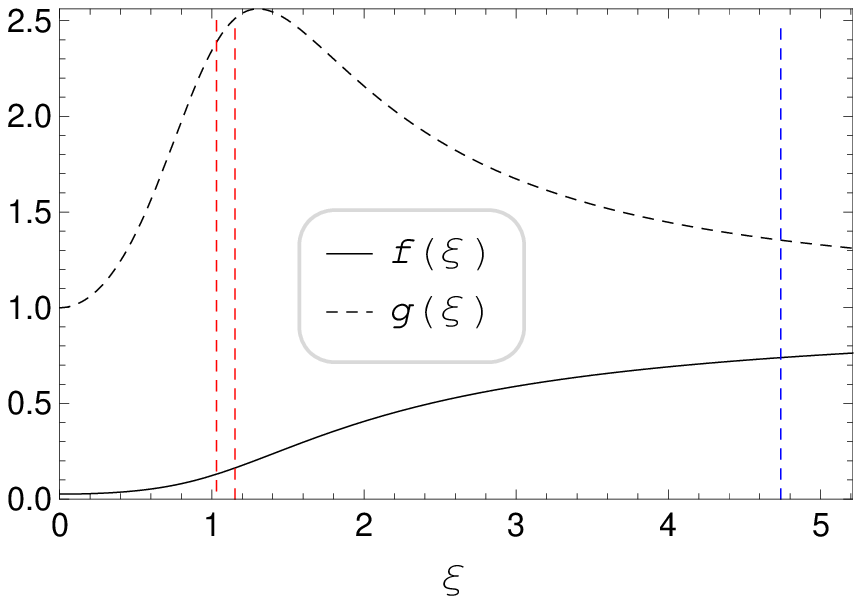}
\includegraphics[width=0.47\linewidth]{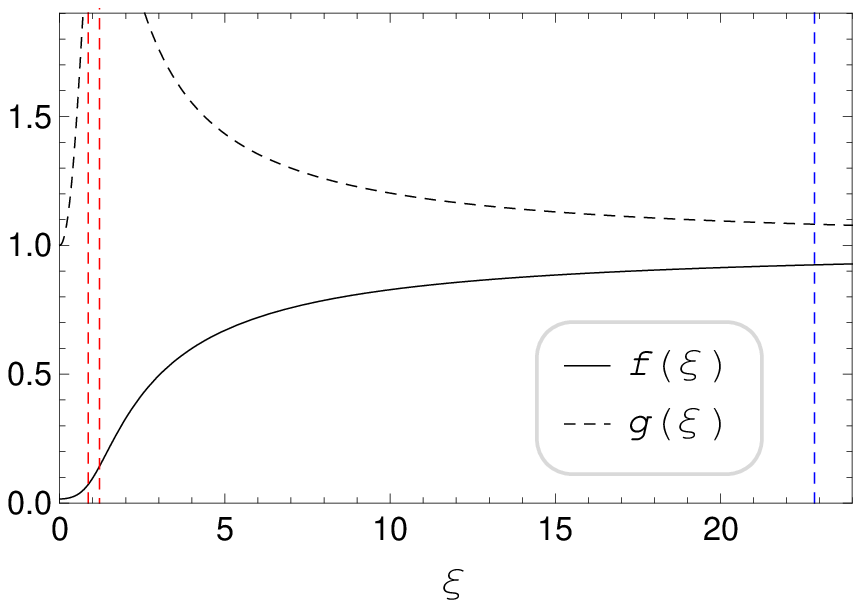}
\includegraphics[width=0.47\linewidth]{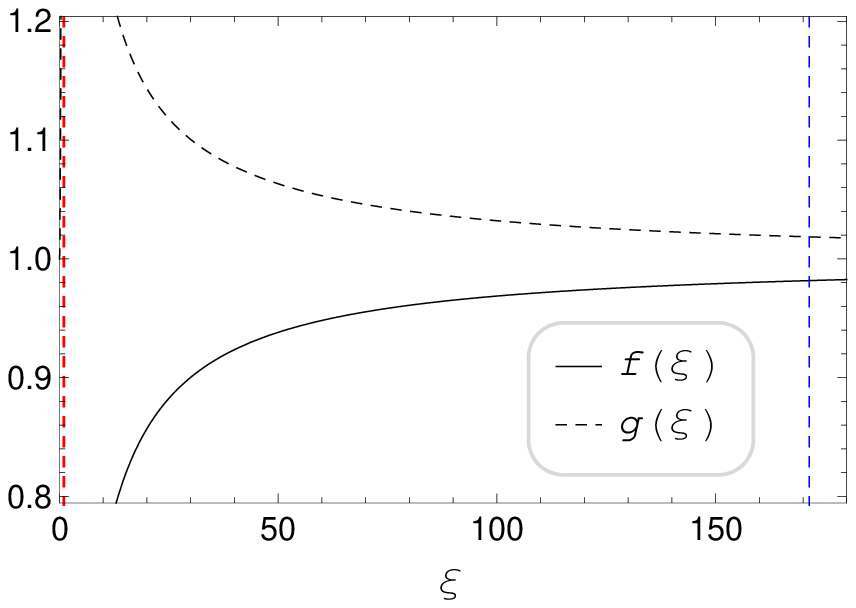}
\includegraphics[width=0.47\linewidth]{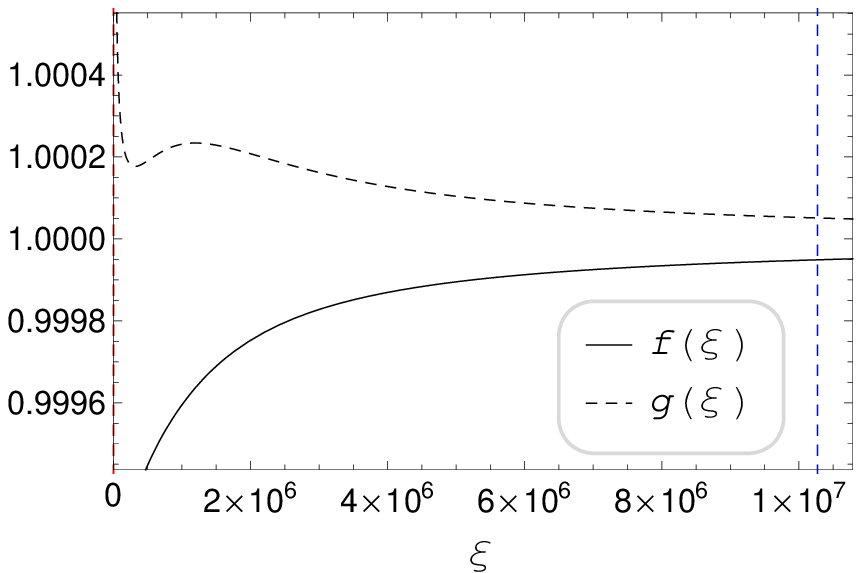}
\end{center}
\caption{\label{fig-lapse-functions} Radial profile of the metric functions $f(\xi)$ and $g(\xi)$ for the typical values of the polytropic parameters. Top panel: (Left panel) $N=2.2$, $\sigma=11/16$, (Right panel) $N=2.7$, $\sigma=27/37$. Bottom panel: (Left panel) $N=3$, $\sigma=18/25$, $\xi_{ph(s)}=0.8583$ and $\xi_{ph(u)}=1.1676$, (Right panel) $N=3.8$, $\sigma=27/34$, $\xi_{ph(s)}=0.6941$ and $\xi_{ph(u)}=1.2015$. Where the vertical, blue and dashed lines represent the radius of the polytrope, $\xi_1$, while the red dashed lines represent the locations of the inner (stable) $\xi_{ph(s)}$ and the outer (unstable) $\xi_{ph(u)}$ circular null geodesics. We can see that extension of the trapping polytropes strongly increases with increasing polytropic index $N$, while the region of strong gravity with extremal values of the metric coefficients and the related location of the trapping zone remains fixed to $\xi \sim 1$, independently of $N$.}
\end{figure*}

For the purpose of our study of the trapping polytropes, it is convenient to use the dimensionless form of the spacetime. We can define the dimensionless line element by the relation $d{\tilde s}=ds/\mathcal{L}$ and dimensionless time by $d\eta = cdt/\mathcal{L}$. The line element of the polytrope spacetime then reads
\bear \label{metric}
d{\tilde s}^2=-f(\xi)d\eta^2+\frac{1}{g(\xi)}d\xi^2+\xi^2d\Omega^2\ ,
\ear
where $d\Omega^2=d\theta^2+\sin^2\theta d\phi^2$. The temporal metric coefficient of the polytrope interior reads
\beq
    f(\xi) = (1+\sigma\theta)^{-2(N+1)} \left\{1-2\sigma(N+1) \frac{v(\xi_{1})}{\xi_{1}} \right\},
\eeq
while the radial metric coefficient of the polytrope interior reads
\beq
    g(\xi) = 1 - 2\sigma(N+1) \frac{v(\xi)}{\xi}.
\eeq
The metric coefficients of the polytrope exterior Schwarzschild geometry (at $\xi > \xi_{1}$) can be expressed in the form
\beq
    f(\xi) = g(\xi) = 1 - 2\sigma(N+1) \frac{v(\xi_{1})}{\xi}.
\eeq
In Fig.~\ref{fig-lapse-functions} the radial profiles of the metric functions $f(\xi)$ and $1/g(\xi)$ are given for typical values of the polytrope parameters. One can see from Fig.~\ref{fig-lapse-functions} that  increasing of the value of the polytropic index $N$ and relativistic parameter $\sigma$ increases the extension of the polytropic sphere, and for sufficiently high values of $N$ there is $\xi_1 >> 1$ for all the allowed trapping polytropic spheres. The extreme increasing of extension parameter $\xi_1$ occurs near the so called critical values of the relativistic parameter $\sigma$, if these values are located at the region of allowed trapping polytropes; at the critical values of the relativistic parameter $\sigma$, existing for $3.3 < N < 5$, extension of the polytropic sphere is unlimited, and for $N>5$ there are no static polytropic spheres at all~\cite{2000NilsonUgla}. On the other hand, extension of the trapping zone given by the position of the unstable circular null geodesic is always restricted to the central region as $\xi_{ph(u)}(N,\sigma) \sim 1$.
\subsection{Circular null geodesics in trapping polytropes}

Four-momentum $p^\mu$ of particles moving along null geodesics satisfies simultaneously the geodesic equation ($\lambda$ is an affine parameter)
\beq
  \frac{\mathrm{D}p^\mu}{\mathrm{d}\lambda}=0,
\eeq
and the normalization condition
\beq\label{normalcond}
  p^\mu p_\mu = 0.
\eeq
Due to the spherical symmetry of the internal spacetime, the motion plane is central, and for a single-particle motion it is reasonable to choose the equatorial plane ($\theta = \pi/2$). The axial symmetry and time independence of the metric coefficients implies conserved energy and axial angular momentum of the particle
\beq
  E = - p_t,\qquad L = p_\phi.
\eeq
The radial component of the geodesic motion, derived using (\ref{normalcond}), reads
\beq\label{radialmot}
   \frac{f(\xi)}{g(\xi)}(p^r)^2 = E^2 - V_\mathrm{eff}(\xi) = E^2 - f(\xi)\frac{L^2}{r^2} ,
\eeq
where $V_\mathrm{eff}(\xi)$ denotes the effective potential of the motion along the null geodesics. For both the internal and external spacetime of the polytrope, the turning points of the radial motion can be thus determined by an effective potential $V_\mathrm{eff}$ containing the axial angular momentum of the motion. The motion is then allowed in the regions where $E^2 \geq V_\mathrm{eff}$ and the equality governs the turning points of the motion.

The local maxima and minima of the effective potential $V_\mathrm{eff}^\mathrm{int}$ directly determine the circular null geodesics and are given by the condition $\mathrm{d}V{}_\mathrm{eff}/\mathrm{d}\xi = 0$, implying the relation
\beq
    \sigma\left[\theta(\xi) + (N+1) \xi \frac{\mathrm{d}\theta(\xi)}{\mathrm{d}\xi}\right] = -1. \label{e21}
\eeq
This condition gives for properly chosen values of the polytrope parameters $N$, $\sigma$ two solutions corresponding to a stable circular null geodesic at $\xi_{ph(s)}$ and unstable circular null geodesic at $\xi_{ph(u)}$, satisfying the relation $\xi_{ph(s)} < \xi_{ph(u)}$. The limiting situation governing coalescence of the circular null geodesics at an inflexion point of the effective potential is determined by the condition $\mathrm{d^2}V{}_\mathrm{eff}/\mathrm{d}\xi^2 = 0$ that takes the form
\bear\label{e22}
&&1 + \sigma\left\{\sigma\theta(\xi)^2 + (N+1)\xi\left[\frac{d\theta(\xi)}{d\xi}\left(4+(2N+1)\sigma\xi\frac{d\theta(\xi)}{d\xi}\right)\right.\right.\nonumber\\
&&\left.\left.+ \xi\frac{d^2\theta(\xi)}{d\xi^2}\right]+\theta(\xi)\left[2+(N+1)\sigma\xi\left(4\frac{d\theta(\xi)}{d\xi}+ \xi\frac{d^2\theta(\xi)}{d\xi^2}\right)\right]\right\}=0\ .
\ear
This condition determines the minimal allowed value of the relativistic parameter $\sigma_{min}(N)$ for allowed values of $N$. For illustration, we give the position of the stable and unstable circular null geodesics in Fig. \ref{fig-lapse-functions} representing the radial profiles of the metric coefficients of the trapping polytropes.

\subsection{Trapping polytropes in $N$-$\sigma$ parameter space}

Using Eqs (\ref{e21}) and (\ref{e22}), we can find that the trapping polytropes can exist for $N>2.138$, and by simultaneous solving of these equations, we can determine the minimal value $\sigma_{min}(N)$ of the relativistic parameter allowing for existence of the trapping polytropes \cite{2017NHS}. The maximal allowed value of the relativistic parameter is given by the causality limit that reads~\cite{1964T-GRP,2016SHN-GRP,2017NHS}
\beq
\sigma_{max} = \frac{N}{N+1}.
\eeq

The numerical analysis presented in~\cite{2017NHS} demonstrates that the trapping region starts to exist for properly selected relativistic parameter $\sigma$, if the polytropic index overcomes the critical minimal value of $N_\mathrm{min} \doteq 2.1378$. The limiting maximal (and simultaneously minimal) allowed value of the relativistic parameter reads $\sigma_\mathrm{max}(N=2.1378) = 0.681$. The results of the numerical analysis are represented in Fig.4 of \cite{2017NHS} that will be used in the following investigation. We can see that the minimal allowed value of the relativistic parameter $\sigma_{min}(N)$ slightly decreases with increasing polytropic index $N$.

\section{Linear gravitational perturbations in polytropic spheres}

The linear gravitational (or scalar and electromagnetic) perturbations were treated for the first time in the case of the Schwarzschild black holes in \cite{1957ReggeWheeler,1970Zerilli}, for the Kerr black holes in \cite{1973Teukolsky}, and for the neutron stars in~\cite{1967ThorneCampolattaro}. These pioneering works were followed by numerous analytical and numerical studies. For perturbations of the spherically symmetric static spacetimes, the axial and polar gravitational perturbations were introduced; their behaviour differs significantly in the case of stellar fluid configurations, as the polar perturbations are efficiently coupled to the fluid, while the axial gravitational perturbations are not coupled to the fluid, being similar to those related to black holes. The most actual reviews of the quasi-normal modes of the perturbations of the black holes and star (fluid) configurations can be found in \cite{1999KokkotasSchmidt,2009BertiCardoso,2011KonoplyaZhidenko}.

In the case of the spherically symmetric black hole spacetimes, and the axial perturbations of the spherically symmetric fluid configurations, the scalar, electromagnetic, and gravitational perturbations can be described by the wave equation \cite{2011KonoplyaZhidenko}. In the polytropic spheres the wave equation takes the following dimensionless form:
\bear\label{weq}
\left[\frac{\partial^2}{\partial \eta^2}-\frac{\partial^2}{\partial \xi_\ast^2}+V_{s\ell}(\xi)\right]\Psi_{s\ell}(\xi,t)=0,
\ear
where $\xi_\ast$ is the so called "tortoise" coordinate given by the relation
\bear\label{tc}
d\xi_\ast=\frac{d\xi}{\sqrt{fg}},
\ear
and the effective potential is given by the relation
\bear\label{veff}
V_{s\ell}(\xi)=f\left[\frac{\ell(\ell+1)}{\xi^2}+\frac{1-s^2}{2\xi f} (fg)'
+8\pi(p_\xi-\rho)\delta_{s2}\right],\nonumber\\
\ear
where the prime denotes a derivative with respect to the dimensionless radial coordinate, $\xi$. In the effective potential~(\ref{veff}), the multi-pole number $\ell$ accepts non-negative integers with the restriction $\ell \geq s$, where $s=0$ and $s=1$ correspond to the scalar and electromagnetic perturbations, respectively, while $s=2$ corresponds to the axial gravitational perturbations; $\delta_{s2}$ is the Kronecker delta that survives only for the gravitational perturbations ($s=2$). The additional terms that occur for the fluid configurations, $p_\xi=T^\xi_\xi$ and $\rho=-T^{\eta}_{\eta}$, correspond to the radial pressure and the energy density of the fluid, respectively, being given by the relations
\bear\label{pressure}
&&\rho=\frac{1-g-\xi g'}{8\pi \xi^2},\\
&&p_\xi=\frac{f(g-1)+\xi gf'}{8\pi \xi^2f}.
\ear
For the axial gravitational perturbations, the effective potential~(\ref{veff}) takes the more compact form
\bear\label{veff-grav}
V(\xi)=f\left[\frac{\ell(\ell+1)}{\xi^2}+\frac{2(g-1)}{\xi^2}-\frac{(fg)'}{2\xi f}\right].
\ear
If we consider, due to the spacetime symmetries, the perturbations in the harmonically time dependent form reflected by the relation
\bear\label{harmonics}
\Psi(\xi,\eta)=\Phi(\xi)e^{-i\omega \eta}\ ,
\ear
then the so called stationary wave equation~(\ref{weq}) takes the form
\bear\label{weq2}
\left[\frac{\partial^2}{\partial\xi_\ast^2}+\omega^2-V(\xi)\right]\Phi(\xi)=0.
\ear
where the frequency is assumed to be a complex number
\bear\label{fre}
      \omega=\omega_r+i\omega_i.
\ear
The real part $\omega_r$ represents frequency of the real oscillations, while the imaginary part $\omega_i$ characterizes their damping or growing rate. The cases of $\omega_i<0$ ($\omega_i>0$) represent the damping (growing) of the oscillations. The stationary wave equation has to be solved under proper boundary conditions that have to be defined at infinity, and the black hole horizon, or the fluid configurations centre. Of course, interesting information is obtained by solving the wave equation approximately at regions related to the local extrema of the effective potential.

\section{Quasinormal long-lived axial modes of gravitational perturbation in trapping polytropes}

The solution of the stationary wave equation with related boundary conditions represents the so called eigenvalue problem, where we have to search for the values of $\omega_{n}$ representing eigenvalues of the frequency of the so called quasi-normal modes giving the solution of the eigenvalue problem. For the effective potentials demonstrating local extreme points (related to the stable and unstable circular null geodesics), as is the case under our consideration, the solution of the stationary wave equation gives the so called bound states, and it is convenient to use the so called Wentzel-Kramers-Brillouin (WKB) approximative technique introduced in quantum mechanics, and the Bohr-Sommerfeld quantization rule that is very effective in semi-analytic treatment of the eigenvalue problem~\cite{2011KonoplyaZhidenko}.

It is well known that there is a correspondence of the null geodesic structure of the spacetime to the character of the oscillatory modes~\cite{Cardoso:PRD:79:2009}. In the black hole spacetimes, where an unstable circular null geodesic occurs, the eikonal limit of the quasi-normal modes is governed by the frequency of the circular null geodesic in the real part of the eigen-frequency of the oscillatory mode, while damping of the mode, determined by the imaginary part of the eigen-frequency of the mode, is represented by the Lyapunov exponent characterizing the time instability of the unstable circular null geodesic~\cite{Cardoso:PRD:79:2009}.

In the spherically symmetric ultra-compact uniform density stars, the stable circular null geodesic occurs inside of the star, while the unstable circular null geodesic is present outside the star, in the Schwarzschild spacetime; both the circular null geodesics determine the region of trapped null geodesics. In this trapping region, long-lived axial gravitational quasi-normal modes occur in the linear-perturbation-regime, as demonstrated in \cite{Cardoso:2007:PRD:77,Cardoso:PRD:2014,Macedo:IJMPD:2015}. It is argued in~\cite{Cardoso:PRD:2014} that in the subsequent non-linear regime, ignited by the long-lived axial gravitational perturbations, strong increase of the spacetime curvature can occur in the trapping zone and consequently it can cause collapse and conversion of the trapping zone into a black hole; due to the fact that the axial gravitational perturbations are not coupled to the fluid of the spherically symmetric uniform density configurations, suppression of these axial gravitational modes through any dissipative phenomena is thus not possible.

\begin{figure*}[h]
\begin{center}
\includegraphics[width=0.47\linewidth]{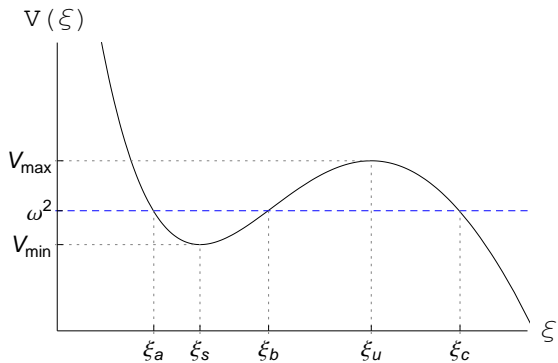}
\end{center}
\caption{\label{fig-potential} Radial profile of the effective potential of the axial gravitational perturbations schematically given for a trapping polytropic sphere. The region of the classical trapping is governed by the local minimum of the potential, $V_{min}$ at $\xi_{s}$, and the local maximum, $V_{max}$ at $\xi_{u}$. For $\omega^2 \in (V_{min},V_{max})$, the turning points $\xi_a$, $\xi_b$, $\xi_c$ are governing the real and imaginary parts of the eigen-frequencies.}
\end{figure*}

Here, we study the quasi-normal axial gravitational modes in the trapping polytropes, giving the eigen-frequencies of the long-lived axial gravitational perturbation modes for typical values of the polytrope parameters $N$ and $\sigma$, covering the polytrope index interval $2.138 < N < 4$. We give a characteristic example of the behaviour of the effective potential governing the long lived quasi-normal modes in the trapping polytropes in Fig.~\ref{fig-potential}.
One can see in Fig.~\ref{fig-potential} that the effective potential for the axial gravitational perturbations~(\ref{veff-grav}) it has one local minimum at $\xi=\xi_{s}$, corresponding to the stable circular null geodesic, and one local maximum at $\xi=\xi_{u}$, corresponding to the unstable circular null geodesic, being located inside the polytropic sphere, $\xi_{s}, \xi_{u} < \xi_{1}$.~\footnote{Recall that in the extremely compact uniform density configurations, the unstable circular null geodesics is located outside the fluid configuration~\cite{2011HS-PNR}.} Let us also stress that the local minimum is located closer to the center of the polytrope than the local maximum, i.e., $\xi_{s}<\xi_{u}$. For values of the frequency parameter restricted by the relation $V_{min} < \omega^2 < V_{max}$, we can obtain three classical turning points located at the radii $\xi_a$, $\xi_b$ and $\xi_c$ that are determined by the value of frequency $\omega^2$. Again, all the radii are located inside the polytropic sphere and satisfy the relations $\xi_{a}<\xi_{s}<\xi_{b}$, and $\xi_{b}<\xi_{u}<\xi_{c}<\xi_{1}$.

We have to stress that both the existence and character of the local minimum and maximum of the effective potential of the axial gravitational perturbations is dependent on the multi-pole number $\ell$, and the trapping polytrope properties -- in dependence on the values of the polytropic parameters $N$ and $\sigma$, the effective potential might not have any local minimum and maximum, if $\ell$ is small enough. Moreover, the boundary conditions restricting the effective potential say that it diverges at the center of the spacetime and tends to zero at infinity
\bear
&&V(\xi)\rightarrow \infty \qquad at \qquad \xi\rightarrow0,\nonumber\\
&&V(\xi)\rightarrow 0 \qquad at \qquad \xi\rightarrow\infty.
\ear

\subsection{Bohr-Sommerfeld approximation}

Generally, the spacetime region where condition $\omega^2\geq V$ is satisfied corresponds to the real energy states where oscillations can occur. In the trapping polytropic spheres, the regions $\xi\in[\xi_a,\xi_b]$ are related to the real energy bound states that can be studied by using the Bohr-Sommerfeld quantization rule~\cite{Andersson:CQG:1993,Cardoso:PRD:2014}. Then the real part of the quasi-normal frequency is given by the Bohr-Sommerfeld rule taking the form as~\cite{Andersson:CQG:1993,Cardoso:PRD:2014}
\bear\label{real}
\int_{\xi_a}^{\xi_b}d\xi_\ast\sqrt{\omega_r^2-V(r)}=\pi\left(n+\frac{1}{2}\right).
\ear

On the other hand, the imaginary part of the energy states, $Im(\omega)$ is related to the tunneling effect of $\omega^2$ through the potential barrier, where $\omega^2\leq V$, i.e, through the region $\xi\in[\xi_b,\xi_c]$. It can be deduced from the shape of the potential that if the value of $\omega^2-V_{min}$ decreases, the real energy state increases and the imaginary part of that decreases because of the exponentially decreasing of probability of the tunneling through wide potential barrier~\cite{Festuccia:ASL:2009,Macedo:IJMPD:2015}. The imaginary part can be given in the form~\cite{Macedo:IJMPD:2015}
\bear\label{imaginary}
\omega_i=-\frac{exp\left[-2i\int_{\xi_b}^{\xi_c}d\xi_\ast\sqrt{\omega_r^2-V}\right]} {4\omega_r\int_{\xi_a}^{\xi_b}d\xi_\ast\frac{1+\sin\left[2\int_{\xi_a}^{\xi}d\xi_\ast\sqrt{\omega_r^2-V}\right]} {\sqrt{\omega_r^2-V}}}\ ,
\ear
The calculations of Eqs.~(\ref{real}) and~(\ref{imaginary}) are performed by substitution $d\xi_\ast=d\xi/\sqrt{fg}$.

As it has been already mentioned, an increase in the value of the multi-pole number $\ell$ causes arising of the local minimum and the local maximum of the effective potential. For sufficiently small values of the multi-pole number, $\ell\approx2$,~\footnote{Sometimes, for sufficiently small values of $\ell$, the effective potential of the trapping polytropes does not have any local minimum or maximum. For example, for the polytropic spheres with the following values of the parameters: $N=2.7$ and $\sigma=27/37$, $N=2.5$ and $\sigma=5/7$, $N=3$ and $\sigma=7/10$, the local minimum and maximum of the effective potential appear for the values of the multi-pole number $\ell\geq3$, $\ell\geq4$, $\ell\geq5$.} the local minimum and maximums of the effective potential are almost at the same order (close each other), and that is why in these cases the probability of tunneling  via the potential barrier, $V_{max}-V_{min}$, is significant and imaginary part of the quasi-normal frequency is high.

In the case of the large multi-pole numbers, $\ell\gg2$, deepness of the effective potential in the bounded region, $V_{max}-V_{min}$, is large and long-lived modes with very small damping rate can occur in such situations. Therefore, in this paper we consider the quasi-normal modes in the eikonal (large $\ell$) limit. In this case, the effective potential of the axial gravitational perturbations of the trapping polytrope spacetime~(\ref{veff-grav}) takes the form~\footnote{Not only the effective potential of the axial gravitational perturbations, but also of the scalar and electromagnetic perturbations can be written in that form in the large $\ell$ limit. However, the correspondence between the null geodesics and quasinormal modes in the eikonal limit is broken in the Einstein-Lovelock theory~\cite{Kon-Stu:2017:PhysLetB:}.}
\bear\label{Null-geo-potential}
V \approx f \frac{\ell^2}{\xi^2}
\ear
that is equivalent to the one governing the propagation along the null geodesics. Then, both the local minimum and local maximum of the effective potential correspond directly to the stable and unstable circular null geodesics (light rings), respectively. Note that because of the independence of the effective potential on the metric function $g(\xi)$, one may expect that in the eikonal regime (for the large multi-pole number $\ell$), the dynamics of the axial gravitational perturbation represented by characteristics such as the state function $\Psi(\xi)$ and the frequency $\omega^2$ does not depend on the metric function $g(\xi)$. However, this is not true since both the metric functions $f(\xi)$ and $g(\xi)$ enter the definition of the tortoise coordinate $\xi_\ast$ in~(\ref{tc}).

By expanding Eqs.~(\ref{real}) and~(\ref{imaginary}) around the minimum of the effective potential, $\xi_s$, and taking only the first terms of the expansion, one arrives at the expression for the real and imaginary part of the quasi-normal frequency in the form~\cite{Cardoso:PRD:2014}
\bear\label{qnm-large-l}
\omega=\Omega_s\ell-iae^{-b\ell}\ ,
\ear
where
\bear\label{Null-circ-frequency}
\Omega_s=\sqrt{\frac{f(\xi_{ph(s)})}{\xi_{ph(s)}^2}}
\ear
is the angular velocity relative to infinity of the motion at the stable circular null geodesic, and $a$ and $b$ are positive constants. One can see that in the large multi-pole number limit, the real part of the quasi-normal frequency is linearly dependent on the multi-pole number $\ell$, while the imaginary part is exponentially suppressed with increasing $\ell$.

In the eikonal limit, the locations of the local extrema depend only on the values of the polytrope parameters $\sigma$, $N$, and tend to finite value. For example, for the case $N=2.2$ and $\sigma=11/16$, the local extrema: $\xi_{s}\approx1.0307$, $\xi_{u}\approx1.1518$ and consequently, $\omega_r\approx0.3498\ell$, while for the case $N=2.7$ and $\sigma=27/37$, the local extrema: $\xi_{s}\approx0.8645$, $\xi_{u}\approx1.2025$ and $\omega_r\approx0.3106\ell$.

We restrict our investigation for simplicity to the fundamental ($n=0$) modes of the quasinormal oscillations. The results obtained in the eikonal limit are compared to the results of the numerical solutions of the evolutionary wave equation~(\ref{weq}), giving the time evolution of the perturbative field $\Psi$ -- for details see subsection~\ref{subsec-numerical}.

\subsection{Numerical results for characteristic trapping polytropes}\label{subsec-numerical}

The direct numerical calculations of the evolutionary wave equation~(\ref{weq}) give interesting and surprising results that are more complex than those obtained for the ultra-compact objects in~\cite{Cardoso:PRD:2014} -- along with the long-lived damping modes that occur for large values of the multipole number $\ell$, corresponding to the eikonal limit, we have found indications of unstable, growing modes for sufficiently small values of $\ell$, depending on the trapping polytrope parameters $N,\sigma$.

The quasi-normal modes of axial gravitational perturbations were calculated in the eikonal limit for typical trapping polytropic spheres, covering the whole region of the allowed values of the polytropic index; we are thus considering also the trapping polytropes with near-critical values of the relativistic parameter $\sigma$. We have chosen four cases of the trapping polytropes for the numerical calculation of the axial gravitational perturbations, concentrating on the calculations of the fundamental, $n=0$, states. The results of the calculations relevant in the eikonal limit are presented in Tabs~\ref{tab1} and~\ref{tab3}.
\begin{table}[ht]
\begin{tabular}{c p{1.5cm} p{1.8cm}}\hline
 $\ell$ ~~~& $\omega_r$ ~~~& $\omega_i$ \\ \hline
60 & 21.1578 & -0.003628 \\
65 & 22.9102 & -0.004523 \\
70 & 24.6616 & -0.004308 \\
75 & 26.4223 & -0.004034 \\
80 & 28.1626 & -0.003752 \\
85 & 29.9127 & -0.003482 \\
90 & 31.6626 & -0.003232 \\
95 & 33.4125 & -0.003010 \\
100& 35.1622 & -0.002803 \\
\hline
\end{tabular}
\quad
\begin{tabular}{c p{1.5cm} p{1.8cm} }\hline
 $\ell$ ~~~& $\omega_r$ ~~~& $\omega_i$  \\ \hline
45 & 14.1379 & -0.000027 \\
50 & 15.7583 & -0.000490 \\
55 & 17.3526 & -0.002628 \\
60 & 18.9341 & -0.002924 \\
65 & 20.5089 & -0.003309 \\
70 & 22.0801 & -0.004404 \\
75 & 23.6493 & -0.006590 \\
80 & 25.2176 & -0.009639 \\
85 & 26.7856 & -0.009228 \\
\hline
\end{tabular}
\caption{Fundamental quasinormal frequencies ($n=0$) of the polytrope with $N=2.2$, $\sigma=11/16$ (left panel) and $N=2.7$, $\sigma=27/37$ (right panel).}\label{tab1}
\end{table}
\begin{table}[ht]
\begin{tabular}{c p{1.5cm} p{1.8cm} }
 \hline
 $\ell$ ~~~& $\omega_r$ ~~~& $\omega_i$ \\ \hline
60 & 17.2472 & -0.000079 \\
65 & 18.7082 & -0.000750 \\
70 & 20.1576 & -0.002090 \\
75 & 21.6002 & -0.002296 \\
80 & 23.0388 & -0.002573 \\
85 & 24.4749 & -0.003225 \\
90 & 25.9094 & -0.004432 \\
95 & 27.3430 & -0.006341 \\
100 & 28.7760 & -0.008365 \\
\hline
\end{tabular}
\quad
\begin{tabular}{c p{1.5cm} p{1.6cm} }
 \hline
 $\ell$ ~~~& $\omega_r$ ~~~& $\omega_i$  \\ \hline
110 & 21.1741 & -2.29e-14   \\
115 & 22.2290 & -1.94e-11   \\
120 & 23.2631 & -2.40e-09   \\
125 & 24.2823 & -6.15e-08   \\
130 & 25.2907 & -6.59e-07   \\
135 & 26.2911 & -7.24e-06   \\
140 & 27.2854 & -2.01e-05   \\
145 & 28.2752 & -1.01e-04   \\
150 & 29.2614 & -1.66e-04   \\
\hline
\end{tabular}
\caption{The same as Tab.~\ref{tab1} but for $N=3$, $\sigma=18/25$ (left panel) $N=3.8$, $\sigma=19/25$ (right panel).}\label{tab3}
\end{table}

We can see that the real part of the frequency, and the magnitude of the negative imaginary part of the frequency, increase with increasing multipole number $\ell$. The most extreme situation for the imaginary part determining the damping properties occurs for the trapping polytrope with $N=3.8$ and $\sigma=19/25$.

Clearly, the existence of the long-lived axial gravitational quasinormal modes has been confirmed in the eikonal limit, but the multipole number has to be very large, $\ell>50$, being strongly dependent on the properties (parameters) of the trapping polytropes.
\begin{figure*}[t]
\begin{center}
\includegraphics[width=0.75\linewidth]{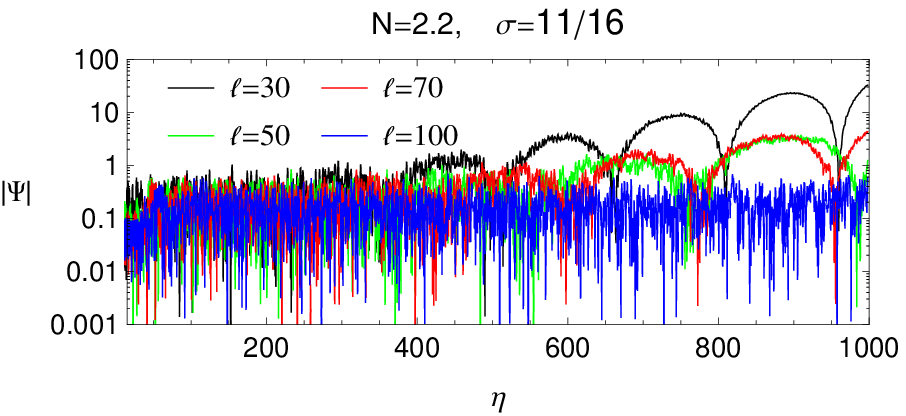}
\includegraphics[width=0.75\linewidth]{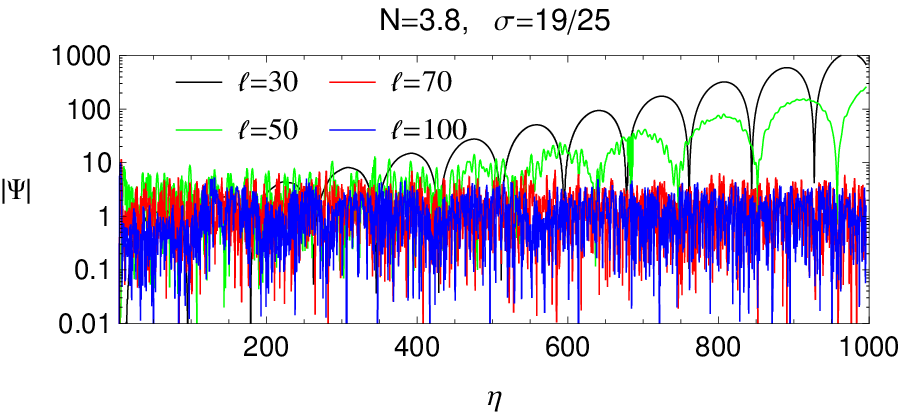}
\end{center}
\caption{\label{fig-time-domain} Time evolution of the axial gravitational perturbations in the trapping polytropic spheres. The sphere with $N=2.2$ is a compact one, while those with $N=3.8$ is extremely extended. For small multipole numbers $\ell$ unstable modes occur, while in the eikonal limit of large $\ell$, the long-lived, damping modes occur.}
\end{figure*}

Direct calculations by numerical code giving the time evolution of the axial gravitational perturbations in the trapping polytropic spheres, represented in Fig.~\ref{fig-time-domain}, show that the Bohr-Sommerfeld quantization rule for solving the eigenfrequency problem in the trapping polytropic spheres is not satisfactorily accurate for small values of the multipole number $l$; on the other hand, these calculations confirm applicability of the eikonal approximation for sufficiently large $\ell$. The value of the sufficiently large multipole number $\ell$ dependes on the parameters of the trapping polytrope.~\footnote{The numerical code has been tested for the standard black holes, giving the standard well known results.} The numerical calculations of the time evolution of the axial gravitational modes governed by Eq. (\ref{weq}) are done for two trapping polytropes: the first one being compact, with $N=2.2$ and $\sigma=11/16$, the second one being extremely extended, with $N=3.8$ and $\sigma=19/25$.

For small values of the multipole number $\ell$, the numerical calculations giving solutions of the time wave equation~(\ref{weq}) clearly indicate existence of growing, unstable axial gravitational modes that would enhance the gravitational instability of the trapping zone. In the large multipole numbers region, the accuracy of the Bohr-Sommerfeld method is very good, as confirmed by the numerical solutions of the evolutionary wave equation~(\ref{weq}), represented in Fig.~\ref{fig-time-domain}, that give long-lived modes predicted by the eikonal limit calculations. On the other hand, as shown in Fig.~\ref{fig-time-domain}, for the trapping polytropic sphere with $N=2.2$ and $\sigma=11/16$, the system is not stable against the axial gravitational perturbations with multipole number $\ell<90$, while for the trapping polytrope with $N=3.8$ and $\sigma=19/25$, the axial modes with multipole number $\ell<70$ are unstable. Therefore, in the $N=2.2, \sigma=11/16$ trapping polytrope case, some of the predictions of the eikonal approximation (namely those for $\ell<90$) are not confirmed by the numerical calculations; in such a situation, for approximate solutions of the wave equation~(\ref{weq2}) with the effective potential~(\ref{veff-grav}), another, more appropriate, semianalytical method should be found, e.g. along the lines presented in \cite{Volkel-Kokotas:2017:CLAQG:}. On the other hand, in the case of the $N=3.8, \sigma=19/25$ trapping polytropes, the eikonal approximation gives the long-lived stable axial gravitational modes for $\ell \geq 110$, strongly exceeding the numerical limit of $\ell \sim 70$. For this reason we cannot exclude that on very long time scales the seemingly stable modes with $70<\ell<100$ could be unstable -- see the extremely low damping rate in the case of the $\ell=110$ mode. We plan more extended study of the growing modes in future study including the influence of the cosmological constant.

\begin{figure*}[t]
\begin{center}
\includegraphics[width=0.45\linewidth]{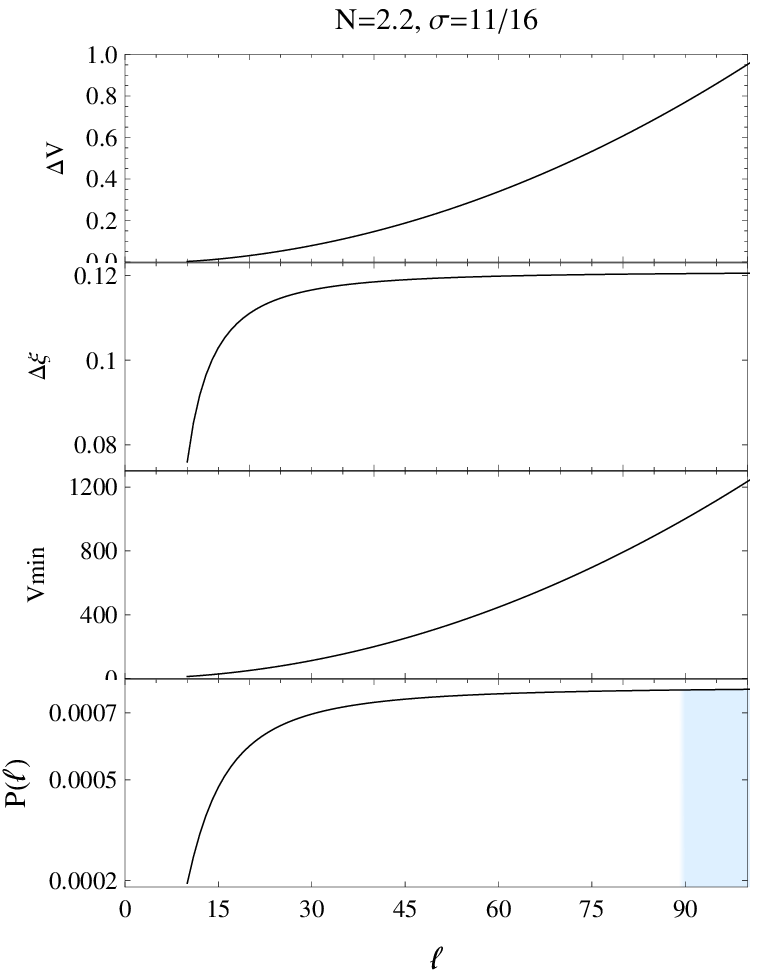}
\includegraphics[width=0.45\linewidth]{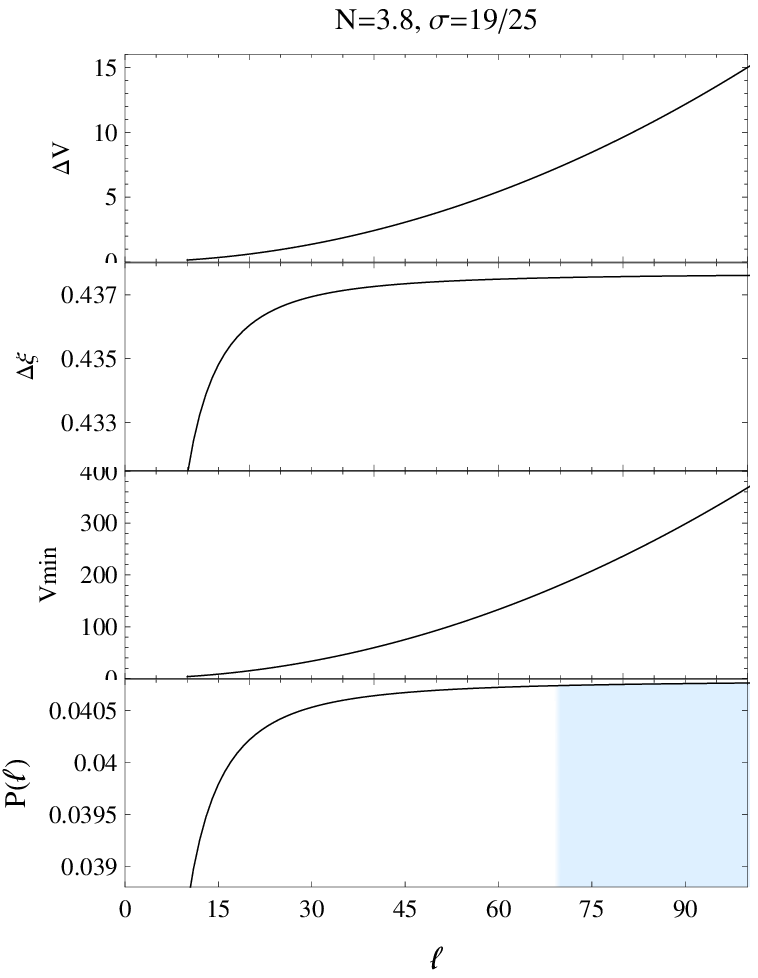}
\end{center}
\caption{\label{fig-poten} Characteristics of the effective potential of the axial gravitational perturbations governing the quasinormal modes in trapping polytropes are given in dependence on the multipole number $\ell$. We present dependences of the difference quantities $\Delta\xi$ and $\Delta V$, the value of $V_{min}$ and the ratio $P(\ell)=\Delta V/V_{min}$. In the $P(\ell)$ dependence we shade the region of applicability of the eikonal approximation supported by the numerical calculations of the time evolution equation~(\ref{weq}).}
\end{figure*}

In order to relate behaviour of the effective potential of the axial gravitational perturbations that occurs in Eq.~(\ref{weq}) to the behaviour of the quasinormal modes, we define, along with value of the minimum of the effective potential $V_{min}(\ell)$, the following crucial characteristics:
\bear
&&\Delta \xi (\ell) = \xi_{u} - \xi_{s},\nonumber\\
&&\Delta V (\ell)= V_{max} - V_{min},\\
&&P(\ell) = \frac{\Delta V}{V_{min}}.\nonumber
\ear
We represent these crucial characteristics of the effective potential in dependence on the multipole number $\ell$ in Fig.~\ref{fig-poten}. We can see that while $\Delta \xi$ does not depend strongly on $\ell$, the values characterizing the extremal points of the effective potential, $V_{min}$ and $\Delta V$, are strongly increasing with the multipole number $\ell$, reaching very large values in the eikonal limit, approving thus its applicability in sufficiently deep and steep gravitational wells, while in shallow gravitational wells the instability occurs in behaviour of the axial modes. We can observe that in case of both considered trapping polytropes, the regions where $\ell$ is sufficiently high, i.e., in the eikonal regime, correspond to $P(\ell)\sim const$ -- here $P(\ell)$ plays crucial role in restrictions of the applicability of the Bohr-Sommerfeld quantization rule for solving the eigenfrequency problem in the trapping polytropic spheres. Thus, in the trapping polytropic spheres the Bohr-Sommerfeld rule (or the WKB method) can have satisfactory accuracy only in the eikonal regime where $P(\ell)\sim const$. As an example, one can see from Fig.~\ref{fig-poten} that for the trapping polytropes with $N=2.2$, $\sigma=11/16$ and $N=3.8$, $\sigma=19/25$, the WKB method~(\ref{qnm-large-l}) can be applied for about $\ell>90$ and $\ell>70$, respectively, in accord with the numerical calculations reflected in Fig.\ref{fig-time-domain}.

We can conclude that for the small values of the multipole number $\ell$, the polytropic spheres containing a trapping region demonstrate existence of unstable, growing axial gravitational perturbations, but with increasing value of the multipole number $\ell$, growing rate of the perturbations decreases and for sufficiently large values of $\ell$, corresponding to the eikonal limit, the axial gravitational modes become stable, but long-lived with small decaying rate, as the modes known from the study of ultracompact objects \cite{Cardoso:PRD:2014}.

\section{Relative extension, mass and compactness of the trapping zone}

Finally, we have to estimate the astrophysical consequences of existence of the long-lived axial gravitational modes in the trapping zone of polytropic spheres, namely the possible consequences of the conversion of the trapping zone into a black hole due to the evolution of this long-lived mode in a subsequent curvature-enhancing non-linear regime, proposed recently in the case of the uniform density ultra-compact stars in \cite{Cardoso:PRD:2014,Macedo:IJMPD:2015}. In order to estimate the fate of the expected black hole creation, we should compare the extension of the trapping zone, and the mass contained within this zone, to the extension and mass of the complete trapping polytropic sphere.

Assume a trapping polytrope with the polytropic index $N$ and relativistic parameter $\sigma$, described by the extension parameter $\xi_{1}$ and the mass parameter $v(\xi_{1})$. Extension of the trapping zone is characterized by the position of the unstable circular null geodesic, $\xi_{ph(u)}$, mass contained within the trapping zone is characterized by the dimensionless parameter $v(\xi_{ph(u)})$. We assume that the mass parameter $v(\xi_{ph(u)})$ also governs mass of the resulting black hole. We introduce two characteristic relative parameters of such a trapping polytrope: the relative extension parameter
\beq
\mathcal{R}_{\xi}(N,\sigma) \equiv \frac{\xi_{ph(u)}}{\xi_{1}} ,
\eeq
and the relative mass parameter
\beq
\mathcal{R}_{v}(N,\sigma) \equiv \frac{v(\xi_{ph(u)})}{v(\xi_{1})} .
\eeq
The mass parameter $\mathcal{R}_{v}$ determines the ratio of the mass of the black hole, created by the gravitational instability and collapse of the trapping zone, to the whole mass of the polytropic sphere. For completeness, we introduce also the relative compactness parameter by the relation
\beq
\mathcal{R}_{C}(N,\sigma) \equiv \frac{C_{max}(N,\sigma)}{C(\xi_1)}= \frac{v(\xi_{max}(N,\sigma))}{v(\xi_1)}\frac{\xi_1}{\xi_{max}(N,\sigma)}
\eeq
where we use, contrary to the previous cases, the maximal compactness $C_{max}(N,\sigma)$ obtained for a given pair of parameters $(N,\sigma)$, related to the global compactness $C(\xi_1; N,\sigma)$.
The maximal compactness occurs at $\xi_{max}$ that is very close to $\xi_{ph(u)}$ for each pair $(N,\sigma)$, and there is always $C_{max}(N,\sigma)<1/3$, as shown in~\cite{2017NHS}. The global compactness strongly decreases for very extended trapping polytropes with $\xi_1\gg1$; recall that, on the other hand, for all the trapping polytropes there is $\xi_{max}\sim\xi_{ph(u)}\sim1$.

\begin{figure*}[t]
\begin{center}
\includegraphics[width=0.55\linewidth]{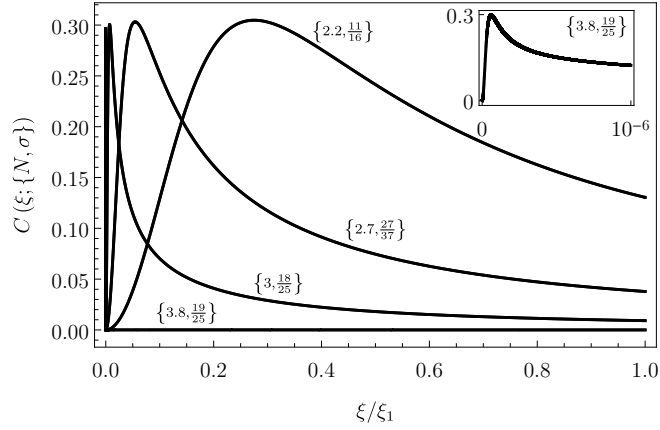}
\end{center}
\caption{\label{fig-7} Radial profiles of the local compactness $C(\xi;N,\sigma)$ of the trapping polytropic spheres. We give the radial profiles for the trapping polytropes with values of the polytrope parameters $(N, \sigma)$ selected in Tables 1 and 2. We use the relative radial coordinate $\xi/\xi_1$, in order to reflect the differences of extension of the polytropic spheres and their trapping zones. In all cases, the local compactness is not reaching the values determining the ultracompact objects, $C>1/3$.}
\end{figure*}

For the pairs of $(N,\sigma)$ selected in Tables~\ref{tab1} and \ref{tab3}, we illustrate the behaviour of the radial profiles of the local compactness $C(\xi)$, in dependence on the relative radial coordinate, $\xi/\xi_1$, in Fig.~\ref{fig-7}. Clearly, the radial profiles of the local compactness expressed in terms of the relative radial coordinate strongly depend on the polytrope parameters, as the maximum of the local compactness is always located at $\xi \sim 1$, but $\xi_1$ can be extremely large for very extended polytropic spheres that can be related to the galactic halos.

\begin{figure*}[t]
\begin{center}
\includegraphics[width=0.45\linewidth]{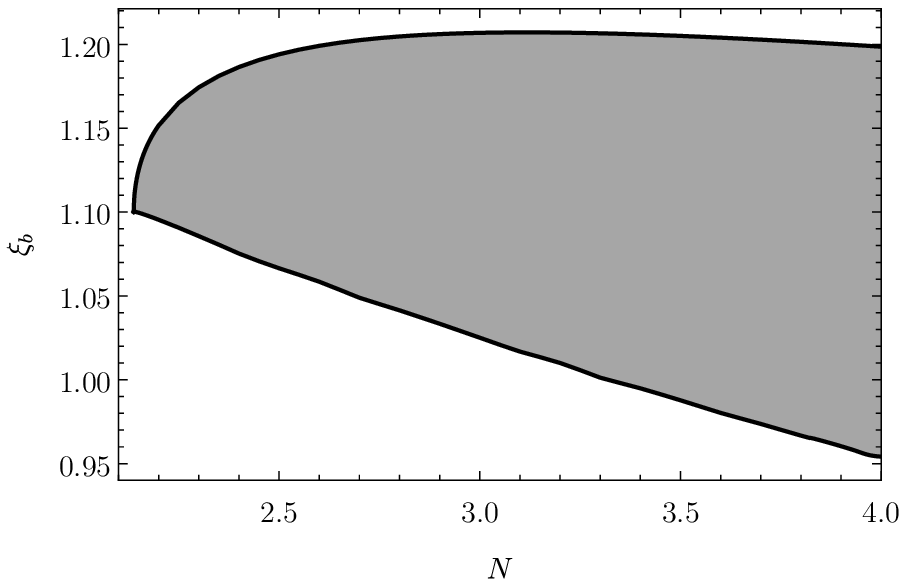}
\includegraphics[width=0.45\linewidth]{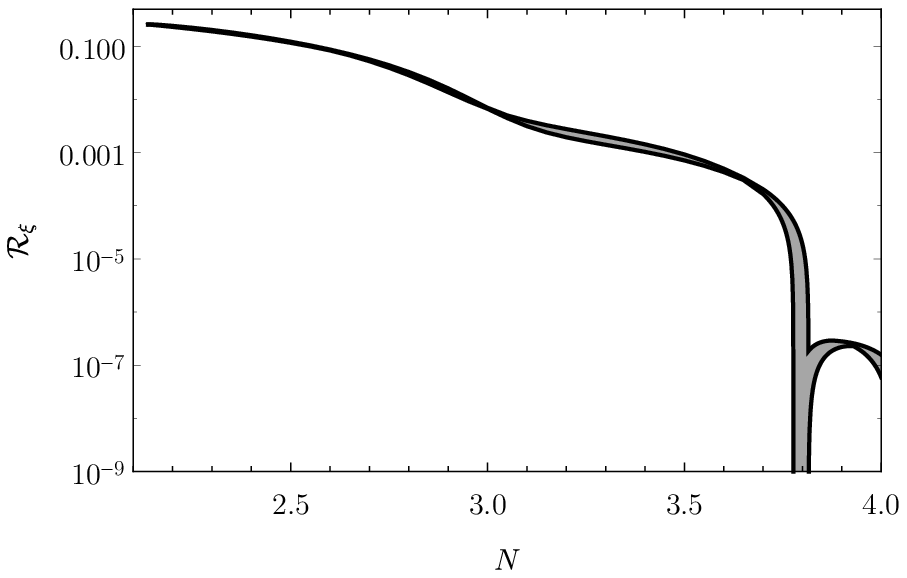}
\end{center}
\caption{\label{fig-4} The relative extension function. Left panel: Dimensionless parameter $\xi_{ph(u)}$ governing the extension of the trapping zone is given as the function of the polytropic index $N$ for the whole region of relativistic parameter $\sigma$ allowing existence of trapping polytropes. Right panel: The relative extension parameter $\mathcal{R}_{\xi}$ giving the relation of the extension of the trapping zone to the extension of the whole polytrope is given as the function of $N$. We have to stress that the minimum (maximum) of the trapping zones do not correspond to the minimal (maximal) allowed value of $\sigma$.}
\end{figure*}

\begin{figure*}[t]
\begin{center}
\includegraphics[width=0.45\linewidth]{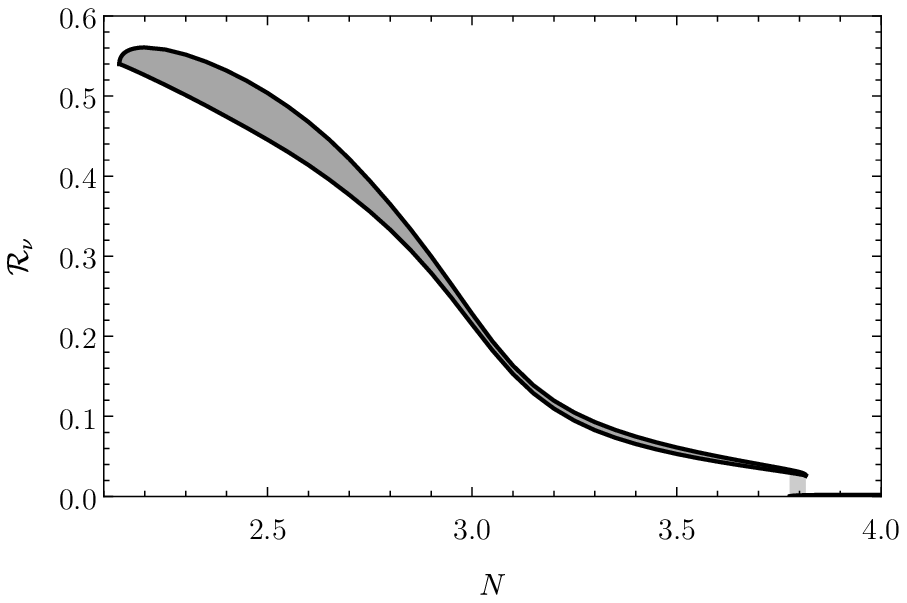}
\includegraphics[width=0.45\linewidth]{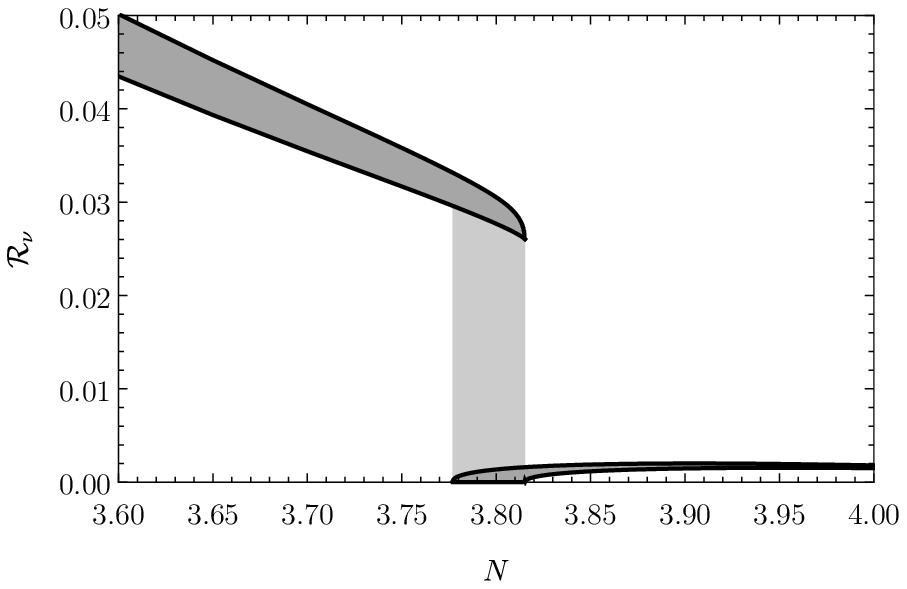}
\end{center}
\caption{\label{fig-5} The relative mass function. The parameter $\mathcal{R}_{v}$, giving the relation of the mass involved in the trapping zone to the whole mass of the trapping polytrope, is given in the left panel as function of the polytropic index $N$ for the whole considered region; its vertical extension corresponds to the whole region of the relativistic parameter allowing the trapping polytropes with given $N$, but the correspondence is not unique, as in the case of the relative extension function. In the right panel we give enlargement of the left panel demonstrating clearly the dichotomy of the trapping polytropes for the values of the polytropic index $N > 3.77$. This dichotomy is related to the appearance of the critical values of the relativistic parameter in the allowance region for the occurrence of the trapping polytropes.}
\end{figure*}

\begin{figure*}[t]
\begin{center}
\includegraphics[width=0.47\linewidth]{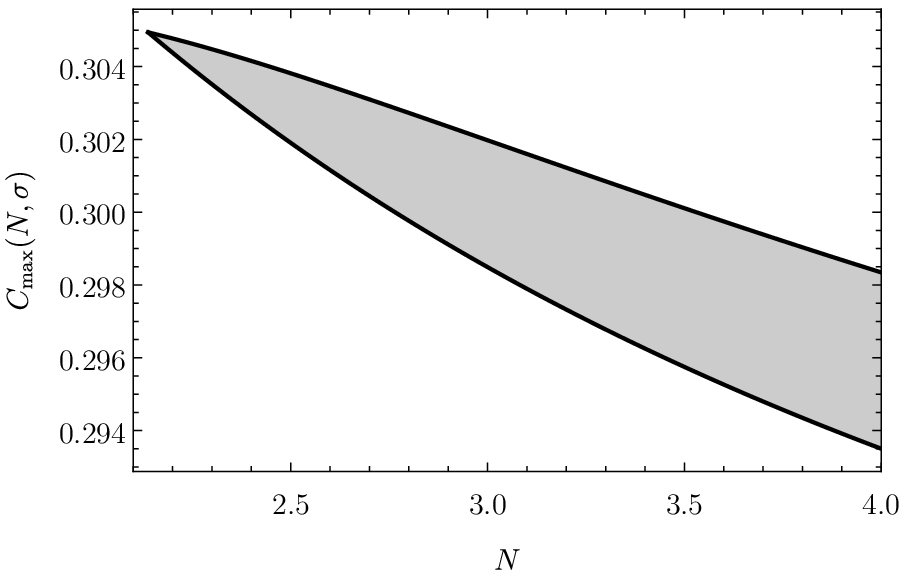}
\includegraphics[width=0.47\linewidth]{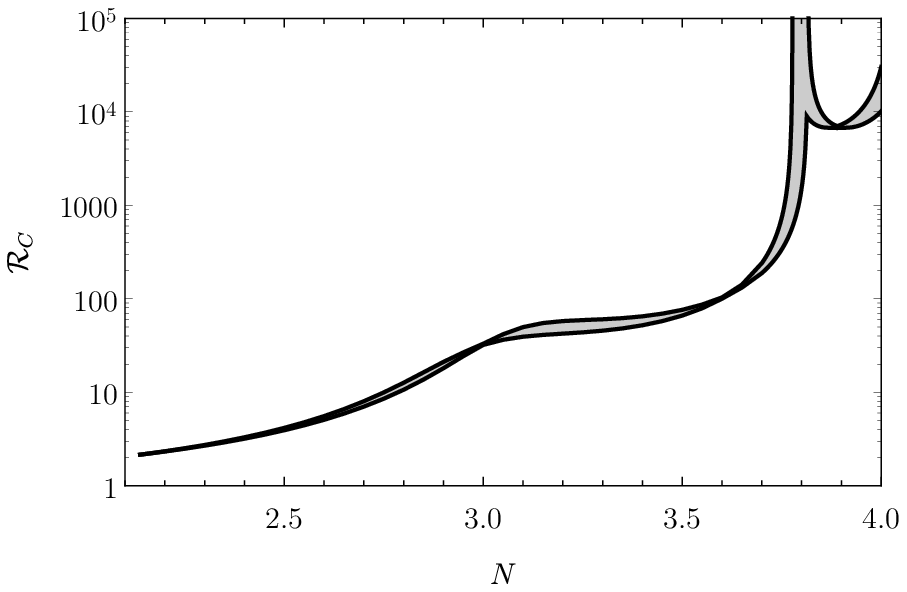}
\end{center}
\caption{\label{fig-8} Relative compactness function of the trapping polytropic spheres. Left panel: The maximal value of the local compactness $C_{max}$ of trapping polytropes, occuring at $\xi_{max} \sim \xi_{ph(u)}$, is given as function of polytrope index $N$, for whole the allowed interval of parameter $\sigma$. In this case the maximal local compactness $C_{max}$ depends monotonically on $\sigma$; the maximal (minimal) value of $C_{max}(N)$ corresponds to maximal (minimal) allowed value of relativistic parameter $\sigma$. Right panel: The relative compactness parameter $\mathcal{R}_{C}(N,\sigma)$ given as a function of parameter $N$. Extension of the region of $\mathcal{R}_{C}$ at a fixed $N$ is determined by extension of the allowed values of parameter $\sigma$. For the relative compactness function we observe at the region of $N>3.78$ a dichotomy behaviour similar to those that occurs for the relative extension parameter. The relative compactness parameter can be in such situations very large, exceeding $10^4$.}
\end{figure*}

The results of the numerical calculations, giving the relative extension function of the trapping zones in the trapping polytropes, $\mathcal{R}_{\xi}(N,\sigma)$, are presented in Fig.~\ref{fig-4}. The numerical results reflecting the relative mass function of the trapping polytropes, $\mathcal{R}_{v}(N,\sigma)$, are presented in Fig.~\ref{fig-5}. The numerical results giving distribution of the maximal compactness $C_{max}(N,\sigma)$ and the relative compactness function $\mathcal{R}_{C}(N,\sigma)$ are presented in Fig.~\ref{fig-8}. In all the figures representing the relative extension, mass and compactness, for a given polytropic index $N$, the dependence on the values of $\sigma$ allowing for existence of the trapping zones is determined by the vertical extension of the areas of $\mathcal{R}_{\xi}(N,\sigma)$, $\mathcal{R}_{v}(N,\sigma)$ and $\mathcal{R}_{C}(N,\sigma)$. However, it must be stressed that all the functions $\mathcal{R}_{\xi}(N)$, $\mathcal{R}_{v}(N)$, $\mathcal{R}_{C}(N,\sigma)$ are not unique functions of the relativistic parameter $\sigma$ for a fixed polytrope index $N$, i.e., the minimum (maximum) of these functions does not correspond to the minimum (maximum) values of $\sigma$ allowing for trapping polytropes.

Character of the relative extension function, $\mathcal{R}_{\xi}(N,\sigma)$, demonstrates relevance of the critical values of the relativistic parameter $\sigma_{crit}(N)$, determining critical behavior of the polytropes that cannot exist for $\sigma=\sigma_{crit}(N)$, and reach extremely large extension in the vicinity of $\sigma_{crit}(N)$, as shown in \cite{2000NilsonUgla,2016SHN-GRP}. The number of the critical values of the relativistic parameter depends on $N$; usually the values of $\sigma_{crit}(N)$ are out of the region of trapping polytropes, but they enter this region for $N \sim 3.8$. Further, we can see in Fig.~\ref{fig-4} that for $N < 3.5$ the trapping polytropes have to be relatively compact, with $\mathcal{R}_{\xi} \leq 10^{-3}$, but in the region of polytropes with $N > 3.5$, the relative extension parameter starts to increase sharply, and for $N > 3.75$, very extended trapping polytropes having $\mathcal{R}_{\xi} < 10^{-7}$ arise that can be extremely extended for $N \sim 3.8$ when effects related to the $\sigma$-criticality are relevant.

For our discussion, the most relevant is the relative mass function $\mathcal{R}_{v}(N)$, reflecting the relative mass content of the trapping zone, as related to the total mass of the polytrope that is crucial for the final fate of the polytropic sphere after expected collapse of the trapping region to a black hole. Clearly, for compact trapping polytropes, with mass of the black hole comparable to the total mass of the polytrope, we could expect that in relatively short time scales the whole polytrope mass could be captured by the created black hole, as predicted in \cite{Macedo:IJMPD:2015}. However, for extremely extended trapping polytropes, where the mass of the black hole created by the gravitational instability of the trapping zone is much smaller than the total mass of the polytrope, and extension of the trapping zone is much smaller in comparison with the polytrope extension, we could expect that various physical phenomena will stabilize the system consisted of the black hole created in the trapping zone and the remaining extended polytropic sphere.

Surprisingly, we have found a double regime for the gravitational instability of the trapping polytropic spheres, clearly reflected in Fig. \ref{fig-5}. This double regime corresponds to two families of trapping polytropic spheres, separated by a jump in the profile of the relative mass distribution function: the compact trapping polytropes, represented by the upper branch, and the extremely extended polytropes, represented by the bottom branch in Fig. \ref{fig-5}.

For values of $N < 3.78$, only compact polytropic spheres can contain the trapping zone, and the mass-distribution function has to satisfy the inequality $\mathcal{R}_{v}(N,\sigma) > 0.03$. If a black hole is created in such a structure, we can expect collapse of the rest of the configuration onto the created black hole if the polytrope extension is comparable with the trapping zone extension, but stabilization of the polytrope could occur, if its extension is much larger than extension of its trapping zone. Notice that in the region of $3.5 < N < 3.81$, the extension of the compact trapping polytropes increases very fast -- see Fig.~\ref{fig-5} where it is demonstrated that the function $\mathcal{R}_{\xi}(N,\sigma)$ sharply decreases from values $\sim 10^{-2}$ to values $\sim 10^{-7}$ corresponding to the extremely extended polytropes, when the remaining part of the trapping polytrope could be stabilized after the creation of the central black hole.

In the second branch, corresponding to the family of very extended trapping polytropic spheres that occurs for $N > 3.78$, there is always $\mathcal{R}_{v}(N,\sigma) \leq 0.001$ and $\mathcal{R}_{\xi}(N,\sigma) < 10^{-7}$, and the stabilization of the trapping polytrope after collapse of its small trapping zone could be realistic. We observe in Fig. \ref{fig-5} that both kinds of these two families of the trapping polytropic spheres can exist for the polytropic index in the region of $3.78 < N < 3.81$.

We can see in Fig.~\ref{fig-8} that the local compactness is not crossing the line $C=1/3$ corresponding to the ultracompact objects. We have to stress that contrary to the mass parameter, $C_{max}$ is monotonically decreasing with $\sigma$ for fixed $N$. The behaviour of $\mathcal{R}_{C}(N,\sigma)$ is, similarly to $R_{\xi}(N,\sigma)$, double-valued in the region $N>3.78$, where it can increase strongly above $\mathcal{R}_C\sim10^4$. For the extremely extended trapping polytropes the relative compactness parameter has to be very large, reaching the values of $\mathcal{R}_C > 10^5$.

In the case of the very extended trapping polytropes, we can expect survival of the polytropic configuration after creation of the central black hole. Such trapping polytropic spheres can be expected to represent galactic halos. If such a halo has a mass $M_{total} \sim 10^{12}M_{\odot}$, we can expect a central supermassive black hole with $M \sim M_{trap} \sim 10^{9}M_{\odot}$ corresponding to typical active galactic nuclei, or smaller, if we consider the polytropes with $\sigma$ approaching $\sigma_{crit}$ from above. Note that the relative mass parameter could be increased by one order and even black holes of mass comparable to $M_{trap} \sim 10^{10}M_{\odot}$ could be acceptable in our scenario, for the polytropes with $\sigma$ approaching $\sigma_{crit}$ from below.

\section{Conclusions}

Recently, existence of standard general relativistic polytropes containing zone of trapped null geodesics has been demonstrated \cite{2016SHN-GRP,2017NHS}. The trapping polytropes can exist, if the polytropic index $N \geq 2.1378$, and the relativistic parameter $\sigma$ is sufficiently high. Of course, it must be lower than the causality limit. The critical value of the relativistic parameter related to the $N = 2.1378$ polytrope reads $\sigma_\mathrm{min}=0.681$. With increasing $N$, the minimal relativistic parameter for trapping  $\sigma_\mathrm{min}$ slightly decreases. For whole range of polytropic indexes, the trapping zone can not exist, if $\sigma < 0.677$.

We have studied the quasi-normal modes of the axial gravitational perturbations in the internal spacetime of the trapping polytropes by using in the WKB approximation the Bohr-Sommerfeld quantum rule. We have explicitly demonstrated existence of the long-lived axial modes that are related to the existence of the trapping zones centered around a stable circular null geodesic in the trapping polytropes. The trapping zones of the trapping polytropes are restricted by the unstable circular null geodesics. The long-lived axial gravitational perturbations could increase in a subsequent non-linear regime.

Existence of the long-lived quasi-normal axial gravitational perturbation modes has been demonstrated in the eikonal approximation of very large $\ell$ and confirmed by solutions of the time-evolution equation. Growing, unstable axial gravitational perturbation modes have been indicated for the smaller values of $\ell$ in the whole variety of the relativistic trapping polytropes with the polytropic index $N>2.138$. Both these cases could lead in the non-linear regime of the gravitational axial perturbations to increase of the spacetime curvature, consequent gravitational collapse of the trapping zone, and possible creation of a black hole. The gravitational instability leading to creation of a black hole in the trapping zone can be realized for all the trapping polytropic spheres, but in two regimes.

For trapping polytropes with $N < 3.5$ the extension and mass of the trapping zone represent a substantial part of the whole polytrope; then the created black hole could capture whole the rest of the polytrope on relatively short timescale. For the trapping polytropes with $3.78 > N > 3.5$, very extended trapping polytropes could contain gravitationally unstable and possibly collapsing very limited innermost trapping zone creating a black hole with mass lower than $0.05$ of the polytrope mass. The most interesting case is related to the special class of the extremely extended trapping polytropes with $N \geq 3.78$, with the trapping zone and the resulting central black hole representing maximally the $10^{-3}$ part of the total polytrope mass, in agreement with observation of the supermassive black hole in the active galactic nuclei.

According to the presented scenario, the supermassive black holes could be created by the gravitational instability of the trapping zones in the early stages of existence of galactic structures when relativistic extremely extended trapping polytropic configurations of dark matter (possibly mixed with baryonic matter) cannot be excluded; a typical supermassive black hole of gravitational mass $M_{BH} \sim 10^9 M_{\odot}$, present in active galactic nuclei, can be related to a typical galaxy halo of mass $M_{halo} \sim 10^{12} M_{\odot}$.

If the extremely extended trapping polytrope model is assumed, extremely large supermassive black holes of mass $M_{BH} \sim 7 \times 10^{10} M_{\odot}$ could be then related to the very large galaxy halos, or halos of galaxy clusters, when mass of the halo reaches values $M_{halo} \sim 10^{14} M_{\odot}$, or can be even by one order higher. However, there is also the possibility to explain such extremely huge central supermassive black hole in a typical galactic halo of mass $M_{halo} \sim 10^{12} M_{\odot}$, if we assume the compact trapping polytropes model with $N \sim 3.7$, having sufficiently large extension to stabilize the black hole-galaxy halo system. Of course, similarly to the models of neutron stars, we could consider galactic structures modelled by several polytropic equations in internal and external regions.

It should be stressed that the case of the extremely extended trapping polytropes, and the related gravitational instabilities of the trapping zones, need a more precise treatment due to the expected influence of the cosmic repulsion represented by the observationally estimated cosmological constant. Although we could expect an insignificant role of the observationally restricted cosmological constant in the behaviour of the long-lived damping or growing axial gravitational modes related to the trapping zones, we expect its substantial role in the restrictions of properties of these trapping polytropes -- such extremely large structures related to large galaxies have to be surely significantly influenced by the cosmological constant, as demonstrated in various similar situations related to toroidal accretion disks, or motion of satellite galaxies \cite{2005S-IRC,Stuchlik:CQG:2009:,Stuchlik:JCAP:2011,Faraoni:JCAP:2015,Faraoni:Galaxy:2016,Arraut:IJMPD:2015,Arrault:Universe:2017}. We expect that relevant answers can be expected due to detailed investigations of the implication of the introductory study of the general relativistic polytropes with the cosmological constant \cite{2016SHN-GRP} where the crucial role of the cosmological constant was explicitly demonstrated. Then we have to give a special attention also to the behaviour of the growing axial gravitational modes in the extremely extended polytropes.

\acknowledgments

The authors have been supported by the Albert Einstein Centre for Gravitation and Astrophysics financed by the Czech Science Agency Grant No.~14-37086G and by the Silesian University in Opava internal grant SGS/14/2016.

\bibliographystyle{JHEP}
\bibliography{gw_polytrope_references}

\end{document}